\documentclass[preprint,5p,twocolumn]{elsarticle}
\pdfoutput=1
\UseRawInputEncoding
\usepackage{geometry}
\usepackage{graphicx}
\usepackage{hyperref}
\usepackage{natbib}
\usepackage{bm}
\begin{document}
\begin{frontmatter}
\title{Detection and Parameter Estimation of Gravitational Waves from Binary Neutron-Star Mergers in Real LIGO Data using Deep Learning}
\author[1]{Plamen G. Krastev\corref{cor1}}
\ead{plamenlrastev@fas.harvard.edu}
\author[2]{Kiranjyot Gill}
\ead{kiranjyot.gill@cfa.harvard.edu}
\author[2,3]{V. Ashley Villar}
\ead{vav2110@columbia.edu}
\author[2]{Edo Berger}
\ead{eberger@cfa.harvard.edu}
\cortext[cor1]{Corresponding author}
\address[1]{Harvard University, Faculty of Arts and Sciences, Research Computing, 38 Oxford Street, Cambridge, MA 02138, USA}
\address[2]{Harvard-Smithsonian Center for Astrophysics, 60 Garden Street, Cambridge, MA 02138, USA}
\address[3]{Simons Junior Fellow, Department of Astronomy, Columbia University, New York, NY 10027, USA}
\begin{abstract}
One of the key challenges of real-time detection and parameter estimation of gravitational waves from compact binary mergers is 
the computational cost of conventional matched-filtering and Bayesian inference approaches. In particular, the application of 
these methods to the full signal parameter space available to the gravitational-wave detectors, and/or real-time parameter 
estimation is computationally prohibitive. On the other hand, rapid detection and inference are critical for prompt 
follow-up of the electromagnetic and astro-particle counterparts accompanying important transients, 
such as binary neutron-star and black-hole neutron-star mergers. Training deep neural networks to identify specific signals 
and learn a computationally efficient representation of the mapping between gravitational-wave signals and their parameters 
allows both detection and inference to be done quickly and reliably, with high sensitivity and accuracy. In this work we apply 
a deep-learning approach to rapidly identify and characterize transient gravitational-wave signals from binary neutron-star 
mergers in {\it real} LIGO data. We show for the first time that artificial neural networks can promptly detect and 
characterize binary neutron star gravitational-wave signals in {\it real} LIGO data, and distinguish them from noise and 
signals from coalescing black-hole binaries. We illustrate this key result by demonstrating that our deep-learning framework 
classifies correctly all gravitational-wave events from the Gravitational-Wave Transient Catalog, GWTC-1 
[{\it Phys. Rev. X} \textbf{9} (2019), 031040]. These results emphasize the importance of using realistic gravitational-wave 
detector data in machine learning approaches, and represent a step towards achieving real-time detection and inference of 
gravitational waves. 
\end{abstract}
\end{frontmatter}

\section{Introduction}

The direct detection of gravitational waves (GWs) by the advanced Laser Interferometer Gravitational-Wave Observatory 
(LIGO) \cite{Abbott:2016blz} confirmed the last remaining prediction of General Relativity and initiated a new 
era of gravitational astrophysics, which enables observations of violent cosmic events that were not previously possible and 
will potentially allow looking directly into the very early history of the universe 
\cite{Abbott:2016blz,TheLIGOScientific:2016src,TheLIGOScientific:2016htt}. During the first (O1) and second (O2)
observing runs the LIGO and VIRGO collaborations reported eleven GW signals from compact binary mergers \cite{LIGOScientific:2018mvr},
which included ten binary black-hole (BBH) events and the first signal from a binary neutron-star (BNS) inspiral, 
GW170817 \cite{TheLIGOScientific:2017qsa}. The observation of GW170817 in both gravitational and electromagnetic (EM) spectra
inaugurated the field of Multi-Messenger Astrophysics (MMA), which uses GWs, EM radiation, cosmic rays, and neutrinos to provide
complimentary information about the astrophysical processes and environments of MMA sources \cite{TheLIGOScientific:2017qsa,GBM:2017lvd}.
The third observing LIGO run (O3) has identified tens of candidate GW signals \cite{Gracedb}, of which thirty nine have been reported 
at the present in the second Gravitational-Wave Transient Catalog, GWTC-2 \cite{Abbott:2020niy}, with four events reported 
previously \cite{LIGOScientific:2020stg,Abbott:2020uma,GW190521,Abbott:2020khf}. With improving GW detector sensitivity 
and with new observatories joining the detector network (KAGRA has joined the detector network \cite{Aasi:2020wya}), 
many more observations, including BBH, BNS and black hole - neutron star (BHNS) events are likely to be detected 
on a regular basis.

Presently, GW data analysis is comprised of two related issues: low-latency searches \footnote{Besides the low-latency searches
there are also "offline" searches, which scan carefully for GW events in detector data.} to enable public release of new 
GW alerts, in part to enable EM follow-up; and GW characterization to estimate GW source parameters and constrain key 
source properties. Currently, the most sensitive low-latency searches targeting GW signals from compact binary mergers 
are based on conventional matched-filtering methods \cite{Gabbard:2017lja,Canton:2014ena}, which typically use large 
banks of pre-calculated waveform templates. Each template is a unique combination of a waveform model and source 
parameters, such as binary component masses and/or spins \cite{Bohe:2016gbl}. These techniques work by taking an inner product between 
the detector data and each template in the bank to generate a signal-to-noise ratio (SNR) time series. This is 
the essence of the matched-filtering approach, which is optimal for identifying signals in stationary, Gaussian 
noise \cite{Maggiore2008}. Because the source parameters are not known in advance, the template bank spans a large 
astronomical parameter space, which makes these approaches computationally expensive and challenging. The computational cost
of matched-filtering methods scales linearly with the number of waveform templates and detectors. Presently, the low-latency
GW surveys target a 4D parameter space (compact binary sources with component masses ($m_1$, $m_2$) and spin-aligned components 
($\bm{\hat{s}_1}$, $\bm{\hat{s}_2}$) on quasi-circular orbits) out of the 9D signal manifold available to the current GW 
detectors (binary component masses ($m_1$, $m_2$) and spins ($\bm{\hat{s}_1}$, $\bm{\hat{s}_2}$) plus the orbital eccentricity $e$) 
\cite{Harry:2016ijz,Huerta:2016rwp}. The computational cost associated with these searches is such that their extension to 
the full 9D parameter space is prohibitive \cite{Huerta:2019rtg}. 

Beyond the detection problem itself, parameter estimation of GWs from compact binaries is formulated as a Bayesian 
inference problem where each likehood evaluation requires generation of the gravitational waveform corresponding to 
a set of compact binary parameters, and computing its noise-weighted correlation with detector data \cite{Chua:2019wwt}. 
Since the waveform generation is the most computationally expensive operation, the GW analysis typically employs 
either faster less accurate waveform models \cite{Hannam:2013oca,Pan:2013rra}, or accelerated surrogates of 
slower more accurate models \cite{Blackman:2015pia}. With current computational resources, obtaining accurate compact 
binary parameters, such as component masses or spins, typically takes hours or days, on the conventional compute grids
typically used by LIGO/Virgo, from the initial detection of the GW event \cite{Chua:2019wwt} which makes the direct 
application of these methods for real-time GW inference unfeasible.

These considerations underline the pressing  need for new, more efficient methods to overcome the limitations of 
the conventional detection and parameter estimation algorithms. Specifically, the need arises for 
approaches enabling real-time detection and inference of GW signals from BNS and BHNS mergers in the full parameter 
space available to current and future GW observatories. Recently, approaches based on deep neural networks have
gained interest in the research community and have been extensively explored as a tool for rapid GW detection and 
inference. Deep Learning (DL) algorithms, a subset of Machine Learning (ML), are highly scalable computational 
techniques with the ability to learn directly from raw data employing artificial neurons arranged in stacked layers, 
named neural networks, and optimization methods based on gradient descent and back-propagation \cite{DL,DL_Book}. 
These techniques, especially with the aid of GPU computing, have proven to be highly successful in tasks such as 
image recognition \cite{DLVis}, natural language processing \cite{DL_NLP}, and recently also emerged as a new tool 
in engineering and scientific applications, alongside traditional High-Performance-Computing (HPC) in the new field 
of Scientific Machine Learning \cite{SciML}. Application of DL approaches in GW astrophysics, specifically 
Convolutional Neural Network (CNN) \cite{CNN} algorithms, were pioneered by George and Huerta \cite{George:2016hay}, 
and Gabbard et al. \cite{Gabbard:2017lja} who demonstrated that CNNs could detect simulated GW signals from BBH 
mergers embedded in Gaussian noise with similar or better performance than that of matched-filtering methods. 
George and Huerta also addressed the GW parameter estimation problem \cite{George:2016hay}. In a subsequent 
work \cite{George:2017pmj} they extended their {\it Deep Filtering} method to detection and characterization 
of GW BBH signals in real advanced LIGO noise with similar performance. Thereafter, DL algorithms have been applied 
in GW astrophysics for detection \cite{Gebhard:2019ldz,Wang:2019zaj,Lin:2020aps,Morales:2020ksv,Xia:2020vem}, 
characterization \cite{Chua:2019wwt,Green:2020dnx} and denoising \cite{Wei:2019zlc} of GW signals 
from BBH mergers.

In our previous work \cite{Krastev:2019koe} we applied, for the first time, a DL approach to detect GW signals from 
BNS mergers, embedded in simulated Gaussian noise, and distinguish them from noise and BBH signals. Our previous 
results demonstrated that deep neural networks can promptly identify weak GW signals from BNS coalescence in 
simulated LIGO noise. 

In this article, we extend our DL framework to detection and parameter estimation of GW signals from BNS mergers 
in {\it real} LIGO noise. We show, for the first time, that DL can be used for both detection and parameter estimation 
of GW signals from BNS mergers embedded in highly non-stationary, non-Gaussian noise. Most importantly, we show 
that ML algorithms, in particular a CNN, can detect {\it real} GW signals from BNS mergers, along with signals 
from coalescing BBHs in a unified DL framework -- our DL approach recovers successfully all GW events from 
GWTC-1 \cite{LIGOScientific:2018mvr}. Furthermore, it is demonstrated that DL can rapidly estimate, with high 
accuracy, the parameters of simulated GW BNS signals embedded in advanced LIGO noise. These results are a step 
towards achieving real-time detection and inference of GWs from BNS (and BHNS) mergers, and thus enabling prompt 
follow-ups of the EM counterparts of these important GW transients.

\section{Methods}

We use two CNNs with similar architecture -- a detection neural network for classification of BBH and BNS GW signals,
and a regression neural network for parameter estimation of GW signals from BNS mergers. The two CNNs differ in the 
output layer -- the detection CNN has a softmax output layer returning the inferred class probabilities, and the 
regression CNN has a linear output layer returning the estimated compact binary parameters, chosen here to be the 
BNS component masses. By adding more neurons to the final layer of the regression CNN, the method can be extended
to estimating additional parameters, such as the BNS sky location, distance, inclination, component spins and/or 
tidal deformabilities. These, and other extensions, are left to future works.

The detection CNN is described in detail in our previous article \cite{Krastev:2019koe}. As before, we distinguish 
between three classes -- BBH signals, BNS signals, and detector noise\footnote{In a subsequent work the detection CNN
will be extended to include also the class of GW signals from BHNS mergers. We do not consider BHNS GW signals in this
article because its main focus is on detection and parameter estimation of GW signals from BNS mergers in real LIGO noise.}. 
Similarly, the training, validation, and testing data sets consist of simulated GW time series where the compact binary 
merger signals (BNS and BBH) are generated using the LIGO Algorithm Library (LALSuite \cite{LALSuite}). Specifically, 
for the BNS signals, we use the TaylorF2 waveform model \cite{Messina:2019uby} and simulate systems with component masses 
in the range 1 to 2 M$_{\odot}$, including also tidal deformation contributions, where the tidal deformability, $\Lambda$, 
is calculated with the APR equation of state (EOS) \cite{Akmal:1998cf}; for calculating $\Lambda$, see e.g., 
Refs. \cite{Hinderer:2009ca,Krastev:2018wtx}. We simulate the BBH signals with the SEOBNRv2 waveform 
model \cite{Purrer:2015tud}, which models the inspiral, merger and ringdown components of the signal, and consider 
systems with component masses of 5 to 50 M$_{\odot}$, with zero spin. The simulated signals are chosen 
to be 10 seconds in duration sampled at 4096 Hz. This choice was made because BNS signals are considerably longer 
and contain typically much higher frequencies than BBH signals. It also reduces the memory requirements during 
the neural network training.  

\begin{figure}[t!]
\centering
\includegraphics[scale=0.5]{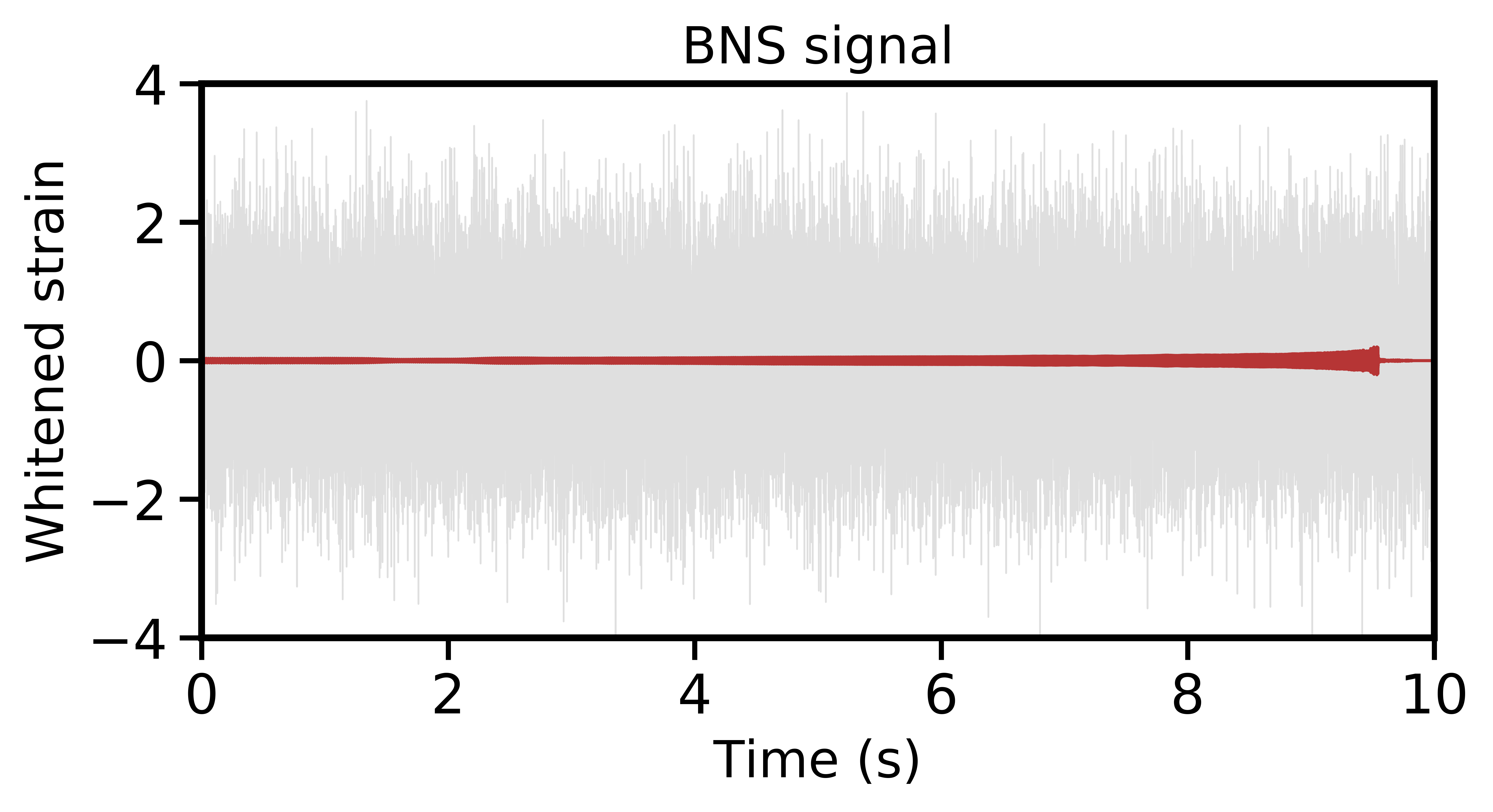}
\includegraphics[scale=0.5]{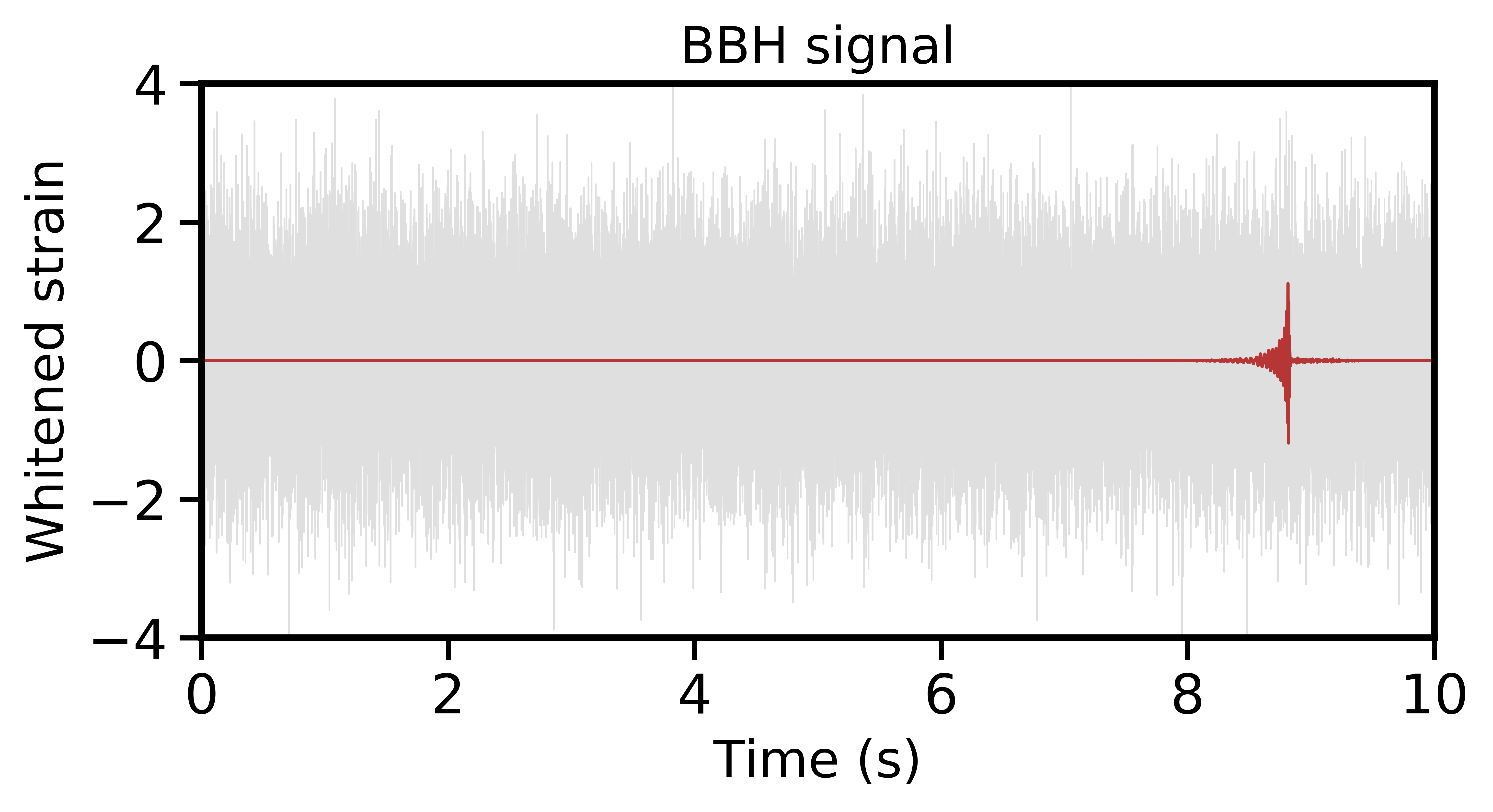}
\caption{{\bf Sample BNS and BBH signals injected in {\it real} LIGO noise.} {\bf (Upper window)} A whitened noise-free time 
series of a binary neutron star gravitational-wave signal with component masses $m_1$=1.4M$_{\odot}$ and $m_2$=1.6M$_{\odot}$ 
and dimensionless tidal deformability  $\Lambda_1=261.9$ and $\Lambda_2=105.5$ (computed with the APR equation of state 
(EOS) \cite{Akmal:1998cf}) with optimal SNR $\rho_{opt} = 8$ (dark red). The values of $\Lambda_1$ and $\Lambda_2$ are
self-consistent with the BNS component masses. The gray time series shows the same gravitational-wave signal with additive 
whitened real LIGO noise of unit variance. This time series is an example of the data sets used to train, validate and test 
the convolutional neural network. {\bf (Lower window)} Same as in the upper window but for a binary black-hole gravitational-wave 
signal with component masses $m_1$=27M$_{\odot}$ and $m_2$=49M$_{\odot}$. ($\Lambda=0$ for black holes.)}\label{fig1}
\end{figure}

We obtained real LIGO data from the LIGO GW Open Science Center (GWOSC) \cite{Abbott:2019ebz}, where we used specifically O2 data 
from the Livingstone detector (L1) sampled at 4096 Hz which does not contain known GW events. Both the data and the
simulated signals are whitened separately with power spectral density (PSD) computed directly from the
raw GW strain data by Welch's method \cite{Welch}. Whitening of data is an operation of rescaling the noise contribution 
at each frequency to have equal power \cite{Gabbard:2017lja}. Because  whitening is a linear procedure, whitening
both parts individually is equivalent to whitening their sum. The waveforms are subsequently shifted such that 
the peak amplitude of each waveform is randomly positioned in the range from 9.65 to 9.95 seconds of the time 
series for the BNS signals, and from 8 to 9.95 seconds for the BBH signals (since BBH signals are considerably 
shorter than BNS signals), to reassure robustness of the network against temporal translations. Different 
realizations of real LIGO noise are superimposed on top of the signals, while the waveform amplitude is scaled 
to achieve a predefined optimal signal-to-noise ratio (SNR), defined as \cite{Gabbard:2017lja}
\begin{equation}
\rho_{opt}^2 = 4\int_{f_{min}}^{\infty}\frac{|\tilde{h}(f)|^2}{S_n(f)}df, \label{eq1}
\end{equation}
where $\tilde{h}(f)$ is the frequency domain representation of the GW strain, $S_n(f)$ is 
the single-sided detector noise PSD and $f_{min}$ is the frequency of the GW signal at the start of
the sample time series. Rescaling the GW waveform is equivalent to moving the source closer or further away 
from the detector, from an astrophysical perspective. Example time series are shown in Fig. \ref{fig1}.

The training sets for the detection CNN consist of 144,000 independent time series with 1/3 containing BNS signal + noise, 
1/3 BBH signal + noise, and 1/3 noise only. The validation data sets consist of 8,000 independent samples containing 
(approximately) equal fractions of each time-series class, and the testing data sets consist of 40,000 samples. We
apply the curriculum learning strategy to train the detection neural network \cite{Shen:2019vep}. This training technique is 
detailed in our previous article \cite{Krastev:2019koe} and is used to optimize network performance and reduce training 
time while retaining performance at high SNR. By starting neural network training at high SNR ($>50$) and then gradually 
increasing the noise in each subsequent training session until the final SNR is in the range between 3 and 20, we found 
that the performance of the detection CNN can be quickly maximized at low SNR (typically after only 10 epochs) while
retaining performance at high SNR.
   
The detection CNN has hyper-parameters similar to those of our original neural network in Ref. 
\cite{Krastev:2019koe}. It consists of 4 convolutional and 4 pooling layers, followed by 2 dense fully connected 
layers. The filter sizes of the convolutional layers are 32, 64, 128 and 256 respectively, and the sizes of the 
dense layers are 128 and 64. We used kernel sizes of 16, 8, 8 and 8 for the convolutional layers and 4 for 
all pooling layers. The first layer corresponds to the input to the neural network which, as before, is a 
one-dimensional time-series vector (of dimension 40,960). At the end, there is an output {\it softmax} layer 
computing the inferred class probabilities. The network architecture was originally optimized for Gaussian
noise with flat PSD, but our experiments indicated that this model also performs optimally with real LIGO
noise with colored PSD. (The architecture of the detection CNN used here is shown in Table 1 of 
Ref. \cite{Krastev:2019koe}.)

The regression CNN used for parameter estimation of GWs from BNS mergers has a similar architecture to the
detection CNN, but instead of an output {\it softmax} layer, it has a linear layer (of size 2) outputting the
estimated masses of the binary components. The architecture of the regression CNN is summarized in Table \ref{tab1}.

\begin{table}[t!]
\begin{center}
\begin{tabular}{r@{\hskip 3mm}lc@{\hskip 3mm}l}
    & Layer            & Array Type & Size  \\
    \hline
    & Input            & Vector     & 40960      \\
 1  & Reshape          & Matrix     & 1 x 40960  \\
 2  & Convolution (1D) & Matrix     & 32 x 40945 \\
 3  & Pooling          & Matrix     & 32 x 10236 \\
 4  & ReLU             & Matrix     & 32 x 10236 \\
 5  & Convolution (1D) & Matrix     & 64 x 10229 \\
 6  & Pooling          & Matrix     & 64 x 2557  \\
 7  & ReLU             & Matrix     & 64 x 2557  \\
 8  & Convolution (1D) & Matrix     & 128 x 2550 \\
 9  & Pooling          & Matrix     & 128 x 637  \\
 10 & ReLU             & Matrix     & 128 x 637  \\
 11 & Convolution (1D) & Matrix     & 256 x 623  \\
 12 & Pooling          & Matrix     & 256 x 155  \\
 13 & ReLU             & Matrix     & 256 x 155  \\
 14 & Flatten          & Vector     & 39680      \\
 15 & Linear Layer     & Vector     & 128        \\
 16 & ReLU             & Vector     & 128        \\
 17 & Linear Layer     & Vector     & 64         \\
 18 & ReLU             & Vector     & 64         \\   
 19 & Linear Layer     & Vector     & 2          \\
    & Output           & Vector     & 2          \\
\hline
\end{tabular}
\end{center}
\caption{{\bf Architecture of the regression CNN.} The architecture of the deep one-dimensional convolutional
network used for parameter estimation of GWs from BNS mergers consist of input layer, followed by 19 hidden layers, 
and output layer returning the estimated binary component masses. The size of the network is about 83 MB.}\label{tab1}
\end{table}

The data sets for our regression CNN are generated from $\sim$400 waveform templates of GWs from BNS mergers by adding
multiple realizations of real LIGO noise and shifting in time. Specifically, for the training data sets we use
$\sim$151,000 data samples with BNS component masses of 1 M$_{\odot}$ to 2 M$_{\odot}$, with $m_1>m_2$. The
validation data sets consist of $\sim$19,000 data samples with intermediate masses in the range of 1.05 to 
1.95 M$_{\odot}$, and the testing data sets consist of $\sim$57,000 data samples with intermediate masses 
of 1.025 to 1.975 M$_{\odot}$.

To build and train the neural networks, we used the Python toolkit Keras (https://keras.io), which 
provides a high-level programming interface to access the TensorFlow \cite{TF} (https://www.tensorflow.org) deep-learning 
library. As in our previous article, we applied the technique of stochastic gradient descent with an adaptive learning rate 
with the ADAM method \cite{ADAM} with the AMSgrad modification \cite{ADAM2}. To train the neural networks, we used an initial 
learning rate of 0.001 and chose a batch size of 1000. For each SNR range the number of training epochs was limited to 100, 
or until the error on the validation data set stopped decreasing. The network training was performed on NVIDIA Tesla V100 GPU 
and the size of the mini-batches was chosen automatically depending on the specifics of the GPU and data sets. In the
detection CNN, we used the sparse categorical cross-entropy loss as a cost function, while in the regression CNN we used
the mean squared error (MSE) loss function.

\begin{figure}[t!]
\centering
\includegraphics[scale=0.5]{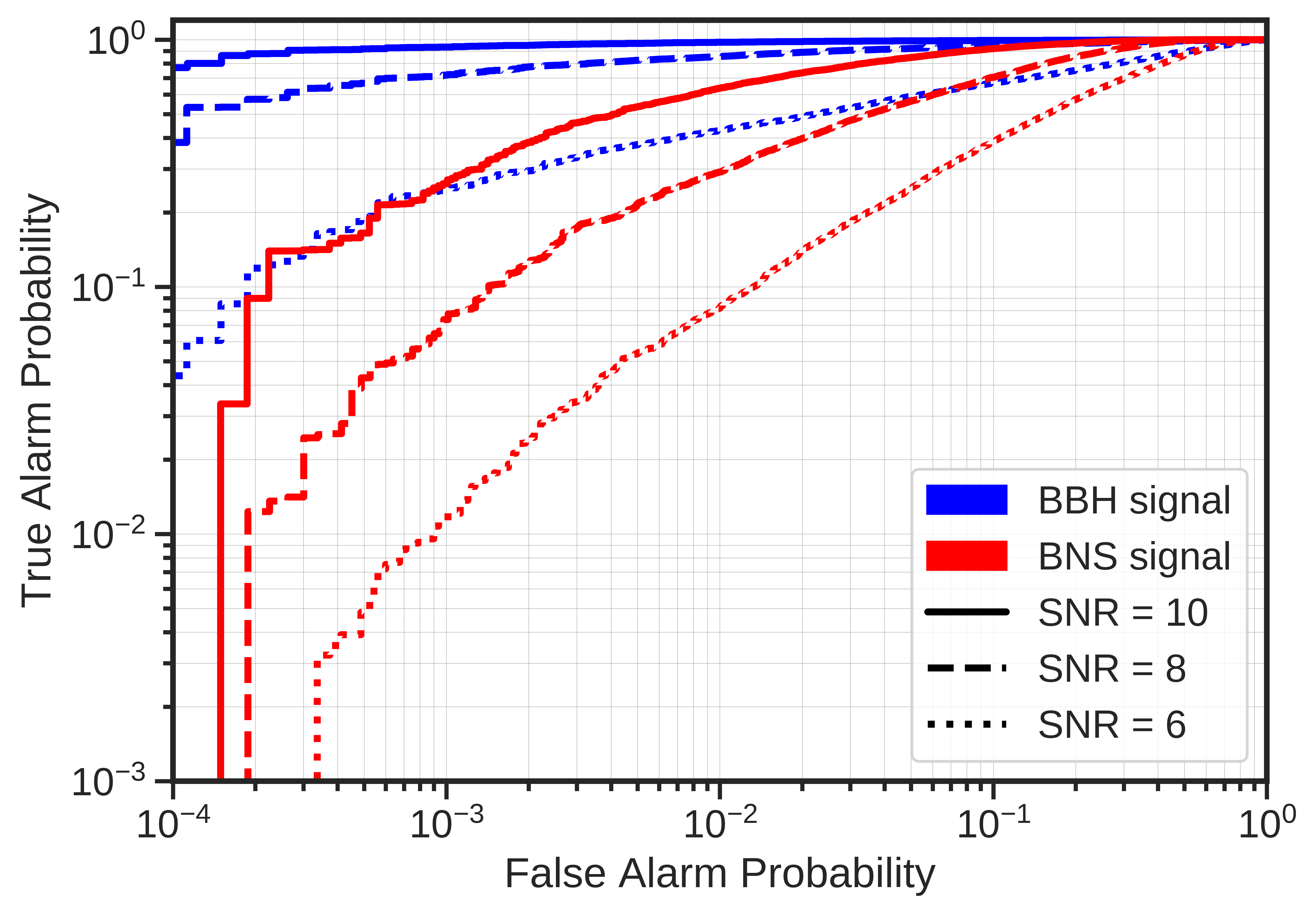}
\includegraphics[scale=0.5]{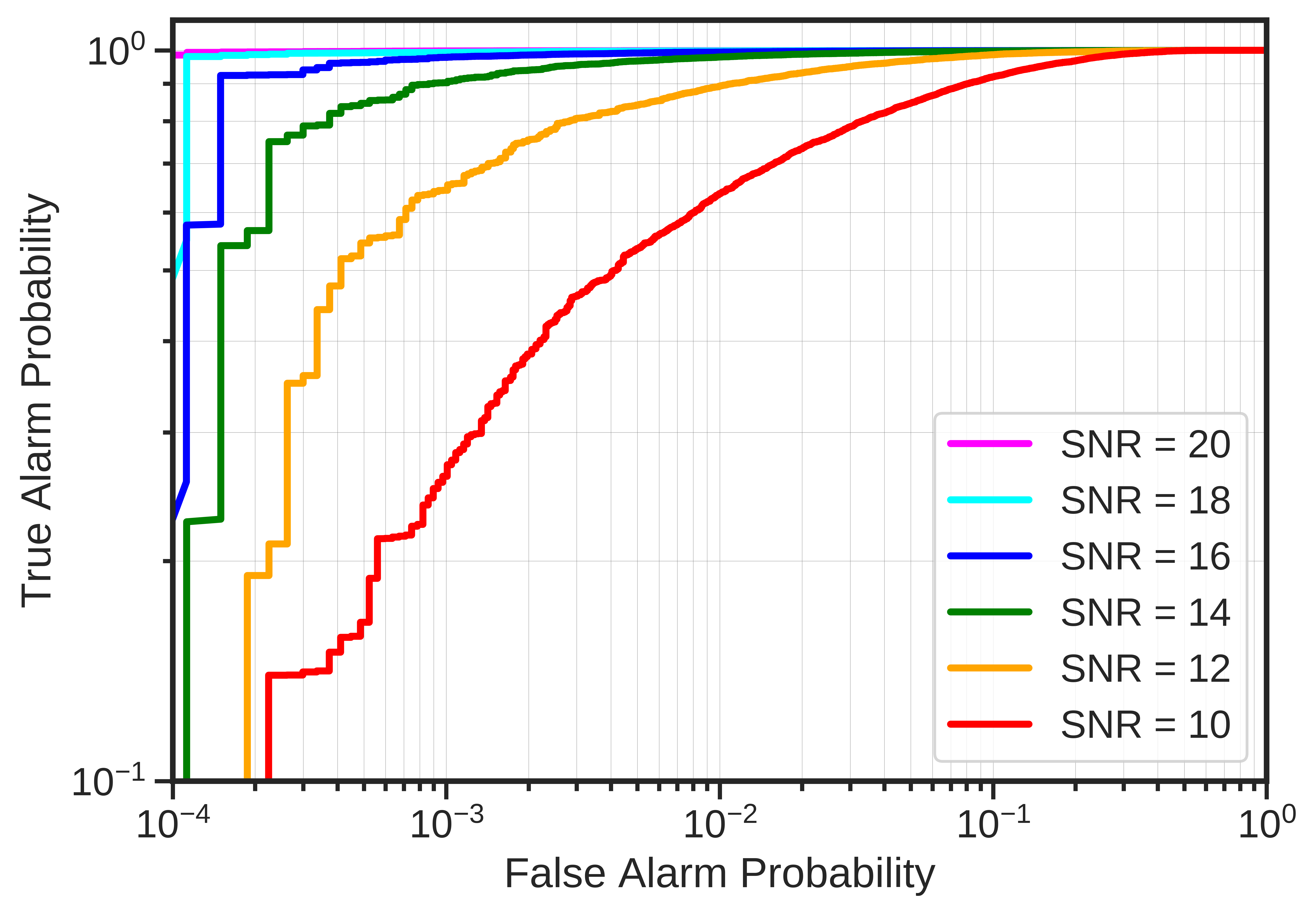}
\caption{{\bf  ROC curves.}  {\bf (Upper window)}  ROC curves for test data sets containing BBH and BNS GW signals with
optimal SNR, $\rho_{opt} = 6, 8, 10$. The true alarm probability is shown versus the false alarm probability 
estimated from the output of the convolutional neural network. {\bf (Lower window)} Same as the upper window 
but only for BNS GW signals with  optimal SNR, $\rho_{opt}$,  in the range from 10 to 20 in steps of 2.}\label{fig2}
\end{figure}

\section{Results}

\subsection{Detection}

Following the same strategy as in our previous work \cite{Krastev:2019koe}, we assess the performance of the
detection neural network by constructing and examining the receiver operator characteristic (ROC) curves for the 
BBH and BNS signal classes, for a given SNR. A ROC curve represents the fraction of signals identified correctly
as their respective class, BBH or BNS (true alarm probability), versus the fraction of samples identified
incorrectly as signals of the particular class (false alarm probability). We calculate the ROC curves with 
the Python scikit-learn library (https://scikit-learn.org), which constructs empirical ROC curves. An empirical 
ROC curve is a plot of the true alarm probability (TAP) versus the false alarm probability (FAP) for all possible 
thresholds, that is, each point on the ROC curve represents a different cut-off value. Thresholds that result
in low FAP also tend to result in low TAP. As the TAP increases, the FAP increases as well. A ranking statistic 
is considered superior to another if at a fixed FAP it reaches a higher TAP (or sensitivity) \cite{Gabbard:2017lja}. 
We varied the optimal SNR from 1 to 20 in integer steps of 1 and the classifier was applied to inputs with 
approximately equal fractions of each GW signal class (Noise, BBH Signal, BNS Signal).

\begin{figure}[t!]
\centering
\includegraphics[scale=0.5]{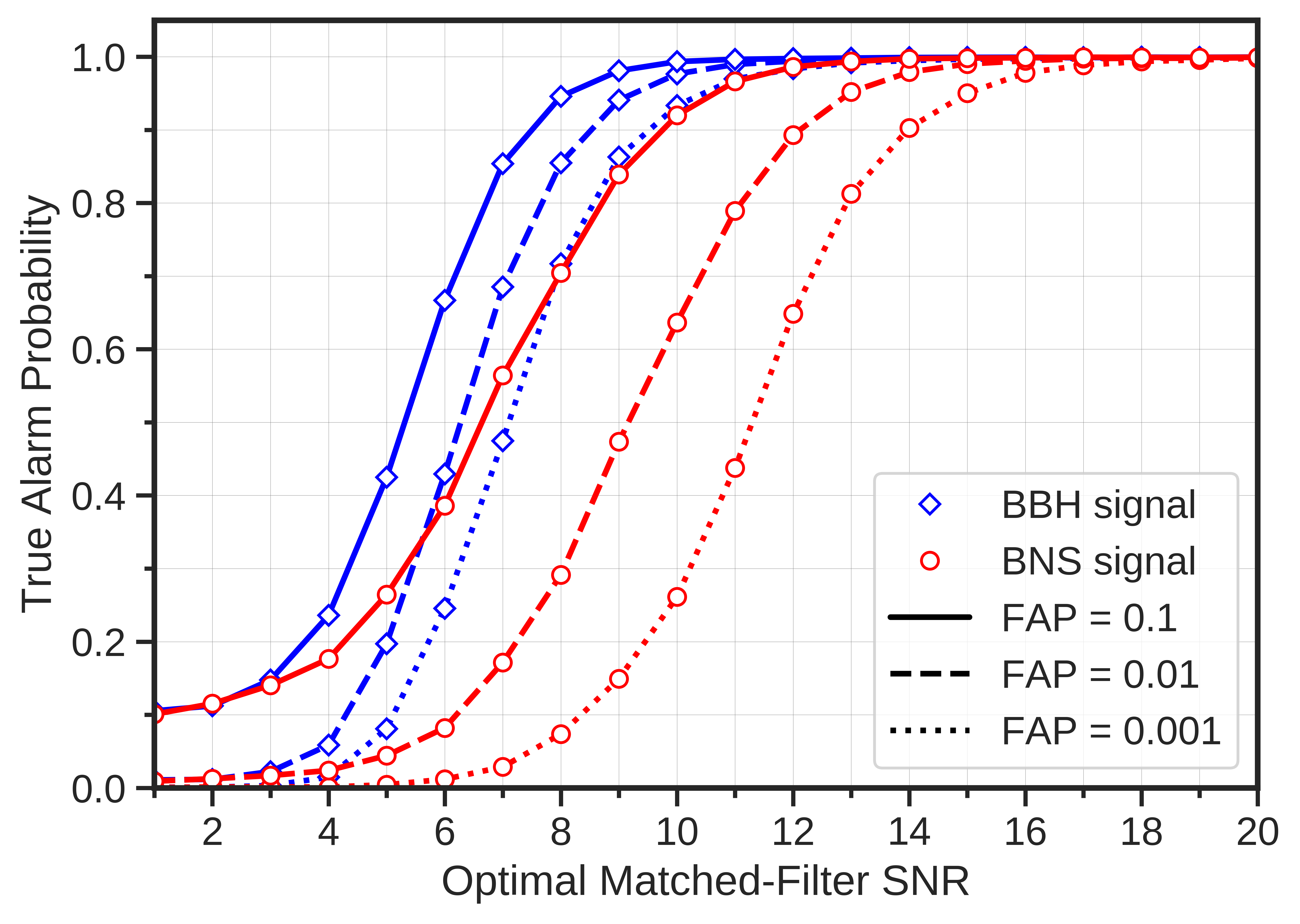}
\caption{{\bf Sensitivity curves illustrating the ability of the neural network to identify BNS and BBH GW signals.}
The true alarm probability is plotted as a function of the optimal SNR for false alarm probabilities $10^{-1}$, $10^{-2}$, 
and $10^{-3}$. The figure shows the sensitivity of detecting GW signals embedded in {\it real} LIGO noise from the test
data set. It is seen that all curves saturate for optimal SNR $\geq$ 18, i.e., all signals are always detected and
correctly classified.}\label{fig3}
\end{figure}

Fig. \ref{fig2} shows the ROC curves calculated for test data sets containing BBH and BNS GW signals. These results
are similar to the corresponding ROC curves in the case of simulated Gaussian noise \cite{Krastev:2019koe}, and
indicate that the neural network is again more sensitive to detecting GWs from BBH than BNS mergers. It is seen that
the CNN achieves a maximal sensitivity for BBH signals with optimal SNR $\rho_{opt} = 10$ for FAP$\geq 10^{-3}$
(Fig. \ref{fig2}, upper window). On the other hand, it reaches a maximal sensitivity for BNS signals with optimal SNR 
$\rho_{opt} = 18$ (Fig. \ref{fig2}, lower window). Note that since the TAP is a function of the FAP, it also
reaches a maximal sensitivity for BNS signals with lower optimal SNR at a higher FAP. For instance, at SNR $\rho_{opt} = 14$
the performance is similar to that at SNR $\rho_{opt} = 18$, but with a FAP of $\sim$0.04.  

The sensitivity of detection of the classifier for different SNR values at a fixed FAP is shown in Fig. \ref{fig3}.
These sensitivity curves are plotted for several representative FAPs ($10^{-1}$, $10^{-2}$, $10^{-3}$) and represent
the ability of the neural network to identify GW signals from BNS and BBH mergers. The lowest FAP used in our analysis
translates to a false alarm rate (FAR) of 0.1\%, or an estimated FAR of $\mathcal{O}(10^3)$ per month\footnote{The 
FAR is estimated from the false alarm probability assuming overlapping GW time series segments of duration 0.3 seconds 
to match the length of the interval within which the peak amplitude is varied. Then, the FAR is rescaled to false 
alarms per month.}. The FAR can be decreased by applying the classifier independently to multiple GW detectors
and enforcing coincidence \cite{Wei:2020ztw}, and  also by checking the consistency of the estimated GW source 
parameters. Furthermore, these false alarms can be quickly eliminated by running conventional matched-filtering 
pipelines with several templates with parameters close to the values predicted by the regression CNN \cite{George:2017pmj}. 
We observe that all curves saturate (at 1) for optimal SNR $\geq$ 18 which implies that all signals, both BNS and BBH, are 
always detected. Note that the sensitivity curves for the BBH signals saturate at a lower SNR of $\sim$ 12. These 
results follow closely our previous findings \cite{Krastev:2019koe}, in the case of Gaussian noise, and demonstrate 
the ability of the detection CNN to identify BBH and BNS GW signals in highly non-stationary, non-Gaussian detector 
noise. The BBH sensitivity curves are also similar to the results reported by George and Huerta \cite{George:2017pmj}.

\begin{figure*}[t!]
\centering
\includegraphics[scale=0.45]{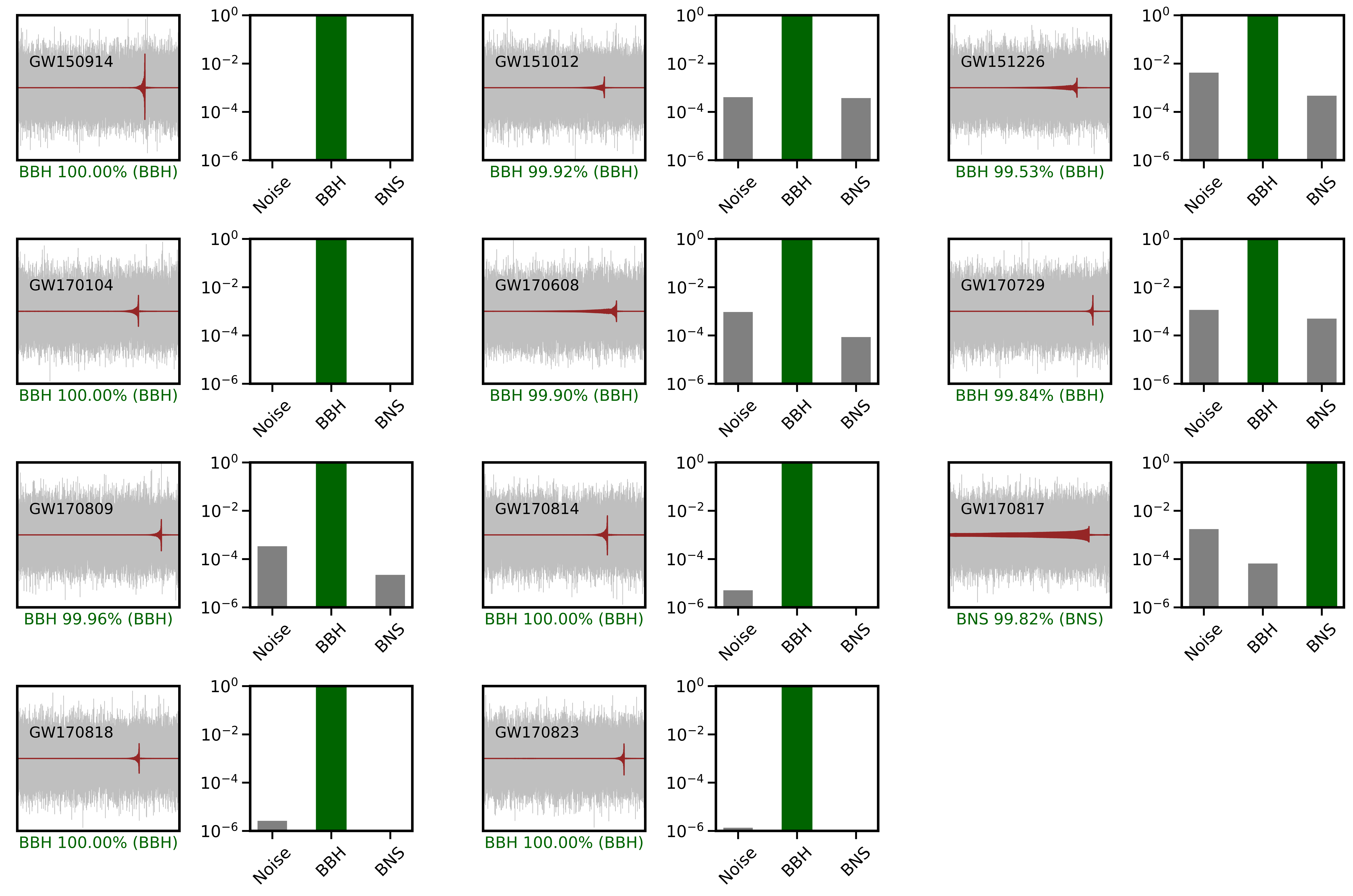}
\caption{{\bf Detection of the GWTC-1 events with a deep neural network trained on simulated GW signals injected
in {\it real} LIGO noise.} The figure shows the probability of detection of each GW event from the Gravitational-Wave 
Transient Catalog, GWTC-1 \cite{LIGOScientific:2018mvr}, and demonstrates the ability of the CNN to identify {\it real} 
BNS and BBH GW events with high confidence. Note the probability of detection is plotted in a log scale.}\label{fig4}
\end{figure*}

To investigate further the sensitivity of the detection neural network, we tested the ability of the classifier
to identify real GW events. We applied the detection CNN to real LIGO data containing all 11 GW events from O1 
and O2 as published in the Gravitational Wave Transient Catalog (GWTC-1) \cite{LIGOScientific:2018mvr}. We obtained
GW time-series data containing the GWTC-1 events from the LIGO GWOSC and prepared the required input for the
detection CNN following the procedure outlined in Section 2. The results from this test are summarized in Fig. \ref{fig4}. 
We find that each GWTC-1 signal is correctly classified as its respective class with a very high probability ($>99.5\%$). 
Most importantly, we show for the first time that a deep neural network can identify {\it real} GW signals from BNS 
mergers and confidently distinguish them from BBH events and detector noise with very high confidence.  

To examine how different noise realizations (simulated or real), used in the  neural network training, impact the 
network's ability to detect real GW signals, we also applied our CNN model trained on GW signals in Gaussian 
noise \cite{Krastev:2019koe} to the GW data from GWTC-1. In this case the classifier identified correctly the 
BNS signal (GW170817), but misclassified 3 BBH events (GW151226, GW170608, and GW170809). In addition, another 2 
BBH events (GW151012 and GW170104) were identified with lower detection probabilities. This implies that when using 
the model trained on GW signals embedded in simulated Gaussian noise, the detection CNN misclassifies about 27\% 
of all real GW events. On the other hand, when using the model trained on GW signals embedded in real LIGO noise, 
it detects and classifies correctly all real GW events. The direct comparison between  the outcomes of these two 
distinct cases (simulated Gaussian noise and real LIGO noise), emphasizes the importance of using  realistic 
GW detector noise in the training procedure of deep neural networks intended for detection of real GW signals.   

\subsection{Parameter estimation}

\begin{figure}[t!]
\centering
\includegraphics[scale=0.5]{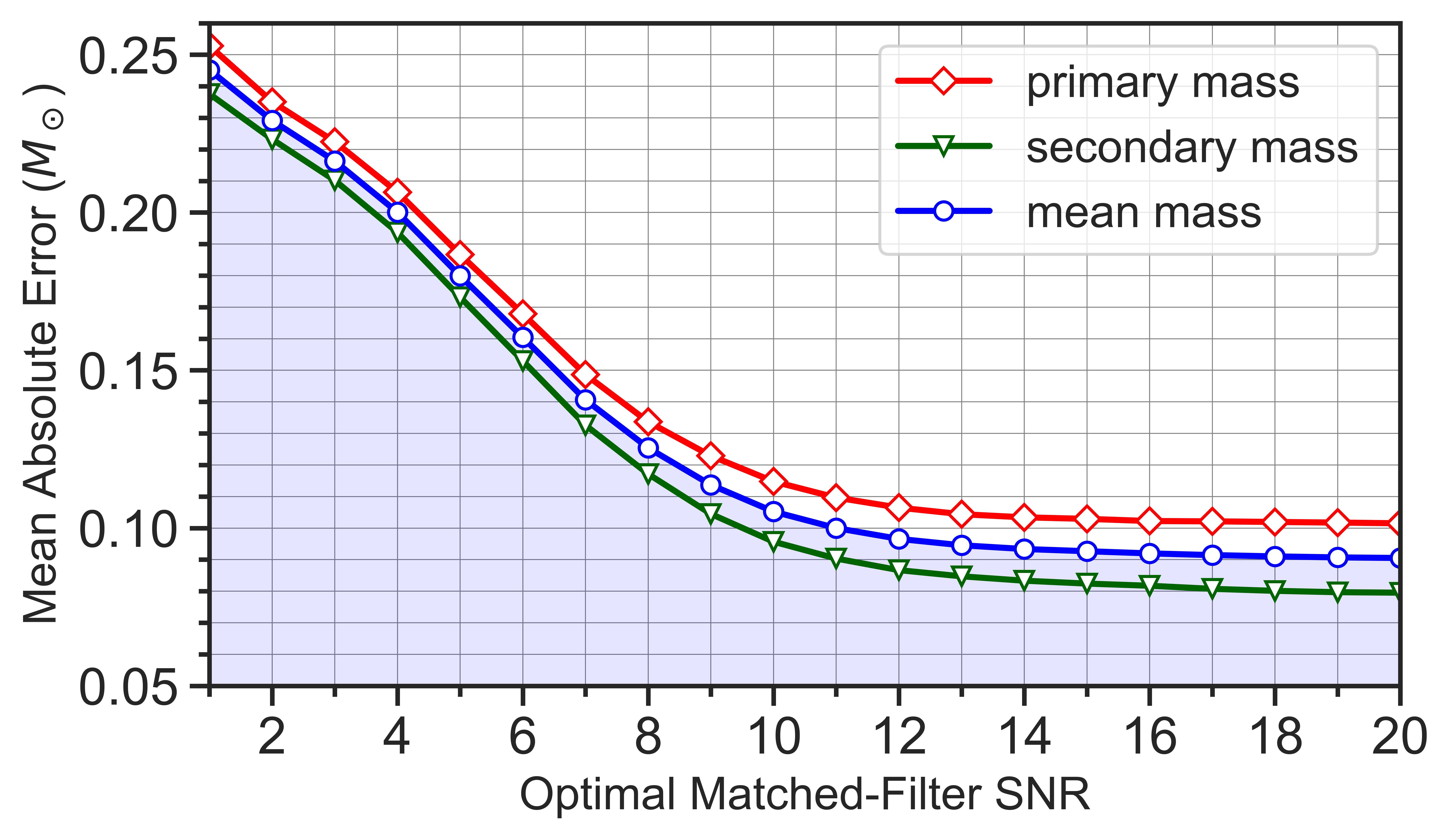}
\caption{{\bf Error in BNS parameter estimation in realistic LIGO noise.} The mean absolute error is shown as a function of 
the optimal matched-filter SNR. The neural network was trained over the whole range of SNR only once, and then the model
was tested on different SNR values without retraining.}\label{fig5}
\end{figure}

\begin{figure}[t!]
\centering
\includegraphics[scale=0.5]{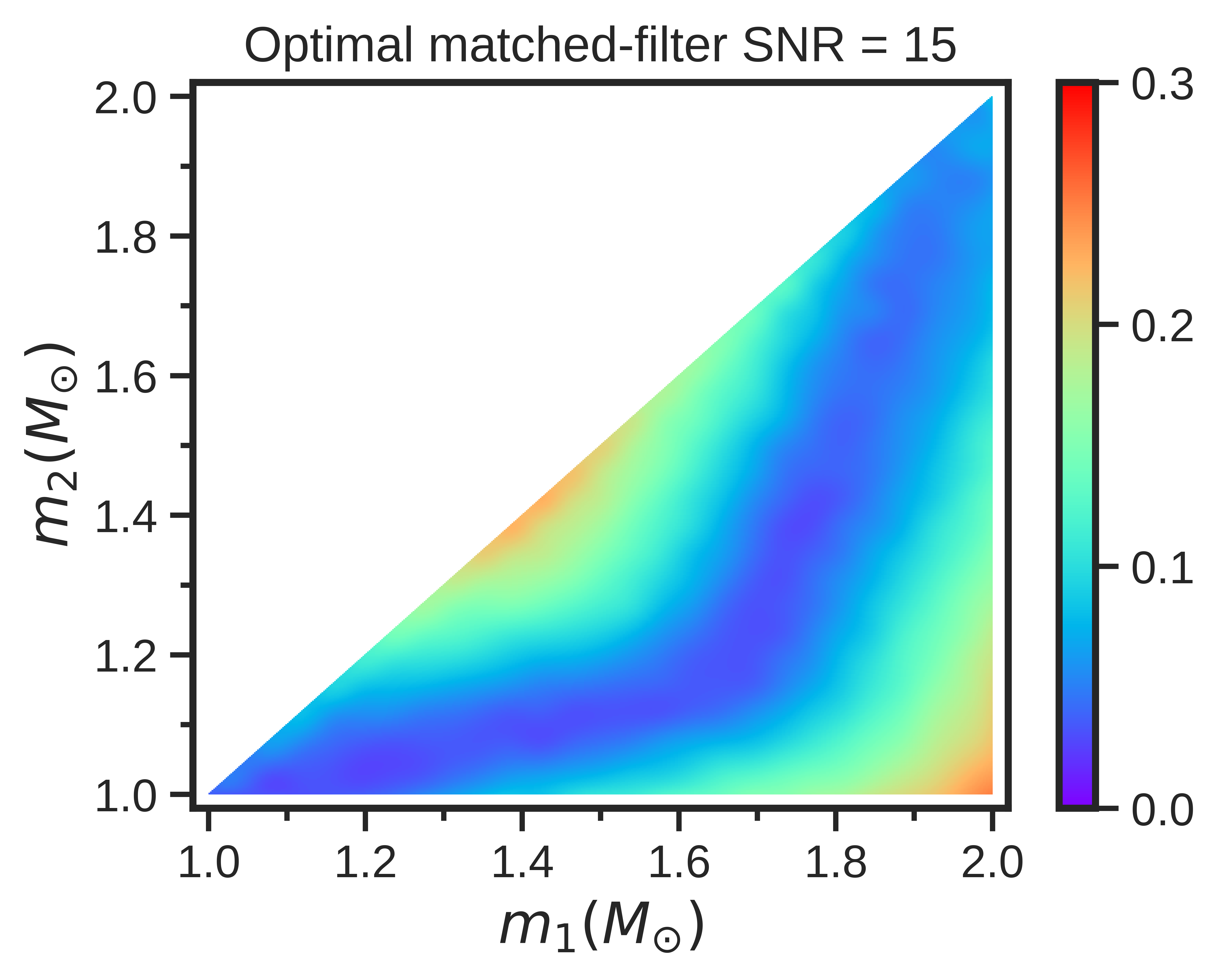}
\caption{{\bf Mean absolute error in predicting the component masses for each template in the
test data set at SNR=15.} This figure shows the mean absolute error in estimating the BNS
component masses for each template in the test data set at a fixed SNR=15.}\label{fig6}
\end{figure}

To assess the performance of our regression neural network, we investigate the variation of the mean absolute error
of the component masses and mean mass of a BNS merger, as a function of masses of the BNS  components versus the SNR;
see Fig.~\ref{fig5}. At a fixed SNR, the mean absolute errors are averaged over all templates in the test data set.
It is seen that the errors decrease with increasing the SNR. The results indicate that for SNR$\geq$11, the regression 
CNN is able to estimate the BNS component masses with a mean absolute error (averaged over all templates and both BNS 
components) smaller than 0.1 M$_{\odot}$, which is about an order of magnitude larger than the spacing
between the templates in the parameter space ($\sim 10^{-2}M_{\odot}$). The predicted BNS component masses are 
deterministic point-wise estimates obtained by extending the methods of Refs. \cite{George:2016hay, George:2017pmj} 
to inference of GWs from BNS mergers. In this respect, the corresponding (mean) absolute errors are also deterministic 
and point-wise, depending only on the testing data sets at each SNR. The point-wise nature of the errors is illustrated 
in Fig.~\ref{fig6}, where the dependence of the error on the BNS component masses for each template in the test data 
set is shown at a fixed SNR=15. The figure displays the error with which the binary components are estimated in each region 
of the parameter space. We find that the error decreases with increasing the SNR in all regions of the covered parameter 
space. Furthermore, for SNR$>$8 the largest errors are clustered in two distinct regions -- one with largest mass ratio, 
and another toward the middle of the parameter space diagonal.  

\begin{figure}[t!]
\centering
\includegraphics[scale=0.5]{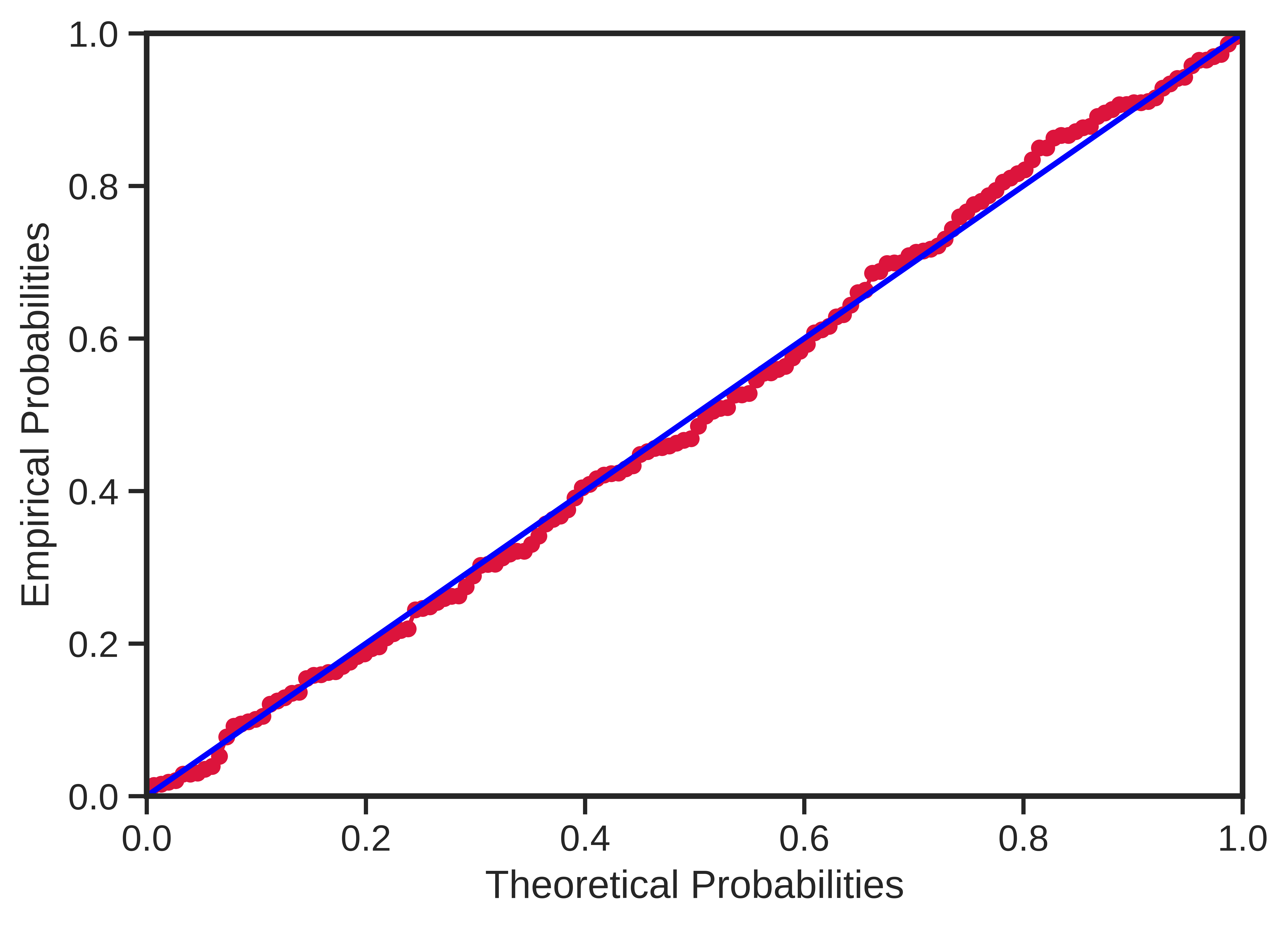}
\caption{{\bf P-P plot of errors in BNS parameter estimation.} The figure shows a P-P plot of the error 
distribution in estimating the primary mass, $m_1$, for waveform parameters $m_1=1.69M_{\odot}$ and 
$m_2=1.5M_{\odot}$ superimposed with 150 different realizations of real LIGO noise at SNR=15. The best
fit is a normal inverse Gaussian distribution with parameters $a=189.63$ and $b=90.40$ (see Eq.~(\ref{eq2})),
and {\it loc} and {\it scale} parameters, -0.11 and 0.31. The errors also followed 
similar distributions in other regions of the parameter space. See text for details.}\label{fig7}
\end{figure}

The error distributions and uncertainties due to superimposing different noise realizations 
were estimated empirically in each region of the parameter space, and it was observed that the
errors followed closely normal inverse Gaussian distributions \cite{norminvgauss} for SNR$>$10.
This allows systematic characterization of error uncertainties. The probability density function 
(PDF) of the normal inverse Gaussian distribution is given by
\begin{equation}
f(x,a,b)=\frac{a K_1(a\sqrt{1+x^2})}{\pi\sqrt{1+x^2}} \exp(\sqrt{a^2-b^2}+bx), \label{eq2}
\end{equation}
where $x$ is a real number, $a$ and $b$ are the tail heaviness and the asymmetry 
parameter with $a>0$ and $|b|\leq a$, and $K_1$ is the modified Bessel function of second kind. The PDF
given by Eq.~(\ref{eq2}) is defined in the "standardized" form, and to shift or/and scale the
distribution, additional {\it loc} and {\it scale} parameters are used to redefine $x$ as
$(x-loc)/scale$. The error distribution was calculated with the {\it statsmodels} Python module
(https://www.statsmodels.org) with the {\it norminvgauss} distribution from the {\it SciPy} library
(https://scipy.org/scipylib). Fig. \ref{fig7} shows a snapshot of the distribution of the errors
incurred in estimating the component masses of a BNS system with component masses $m_1=1.69M_{\odot}$
and $m_2=1.5M_{\odot}$. The errors in other region of the parameter space followed similar distributions.

Here we emphasize that the predicted BNS component masses are deterministic, point-wise estimates, where the
corresponding errors and error uncertainties are based solely on the test data sets at each SNR. They are computed 
extending the methods of Refs. \cite{George:2016hay, George:2017pmj} to GW BNS signals, and should be considered a 
"proof-of-concept" study of parameter estimation of GW signals from BNS mergers in real LIGO noise. In particular, 
the error uncertainties take into account only the errors incurred in predicting the component masses due to the use 
of different noise realizations in the waveform template generation process. Evaluating and understanding the error 
uncertainties can be further improved by recasting the parameter estimation problem into a probabilistic framework.
This can be achieved, for instance, by implementing Bayesian neural networks to perform the GW inference task. Instead 
of having deterministic values, the wights of these networks are characterized by probabilistic distributions by placing 
a prior over the network weights \cite{PerreaultLevasseur:2017ltk}. This approach will be investigated in subsequent works.  

\subsection{Effect of glitches}

\begin{figure*}[t!]
\centering
\includegraphics[scale=0.27]{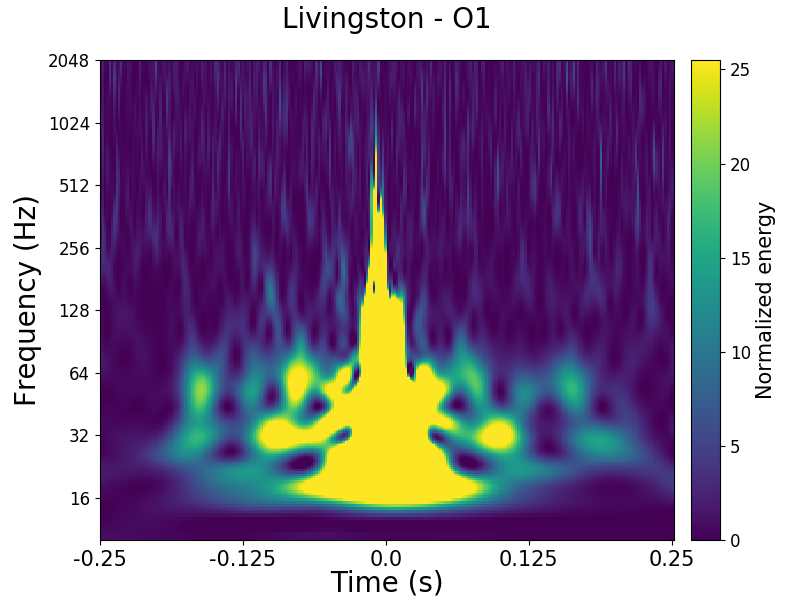}
\includegraphics[scale=0.27]{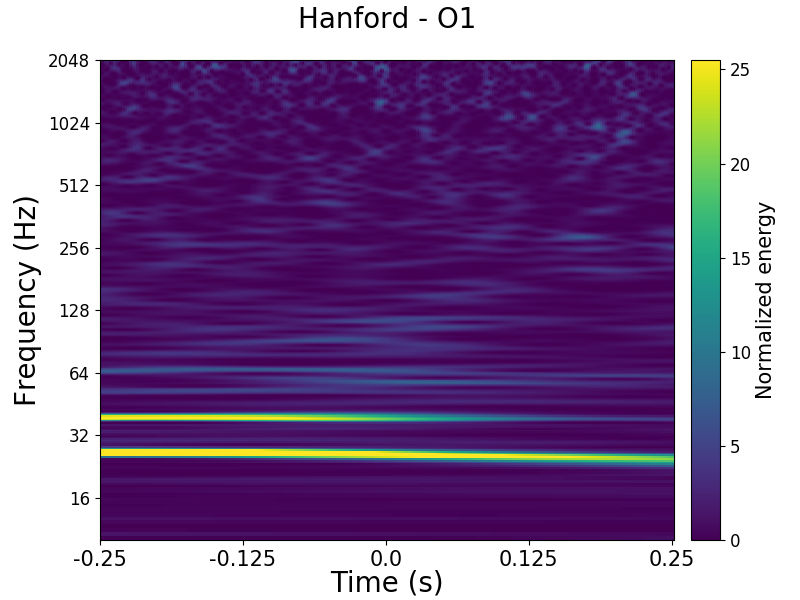}
\includegraphics[scale=0.27]{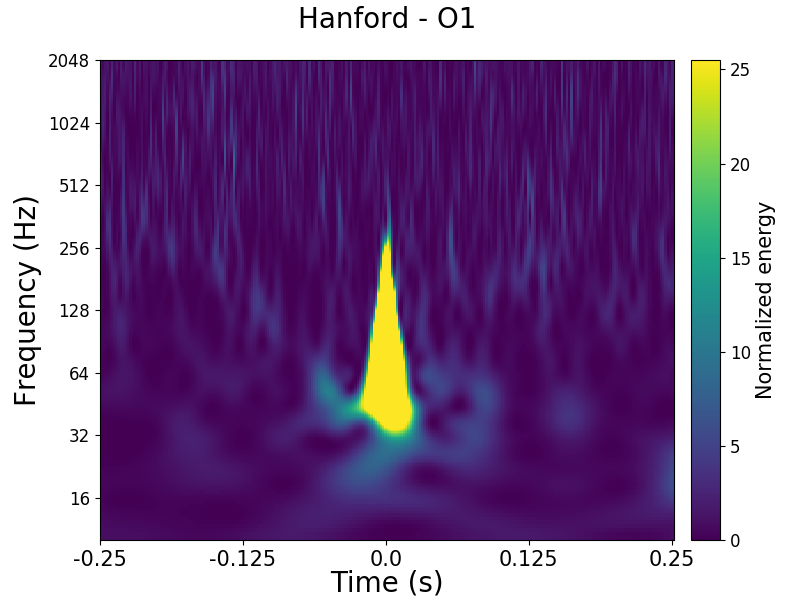}
\caption{{\bf Time-frequency maps of example members of the glitch classes from {\it Gravity Spy} \cite{GravitySpy}.} 
The figure shows normalized time-frequency power maps of example GW time-series for the Koi Fish (left panel), 
Scattered Light (middle panel), and Blip (right panel) glitch classes. We obtained sample images from the Gravity 
Spy public data set \cite{GravitySpyDataSet}. Note that the detection CNN requires as input whitened GW time-series 
templates prepared following the procedure discussed in the Methods section, and these examples solely illustrate 
the morphologies of the three glitch classes considered in this study.}\label{fig8}
\end{figure*}

To examine the effect of transient disturbances on the detection CNN, we performed experiments with a subset of 
realistic glitches from the {\it Gravity Spy} project \cite{GravitySpy}. Specifically, we experimented with three 
glitch classes - Koi Fish, Scattered Light, and Blip. Example members of these glitch classes are shown in Fig.~\ref{fig8}. 
We obtained time-series data containing the glitch events from the LIGO GWOSC and prepared the required input for 
the detection CNN following the procedure discussed in the Methods section.

Without additional retraining, the neural network misclassified $\sim$35\% of the glitches as BBH and BNS signals. 
We next injected $\sim$450 templates from each glitch class (Koi Fish, Scattered Light, and Blip) into the 
training procedure and we found that the detection CNN was able to identify correctly all glitches from our test
data set. During the training phase we grouped the glitches from the three classes together, labeled as "noise", 
and retrained the neural network starting with the model trained on BBH, BNS, and noise samples. Then we used 20 
samples from each glitch class for testing \footnote{The Gravity Spy data set contains also 60 glitches from the 
Chirp class which are "hardware injections" of simulated GW BBH signals physically added to the detectors for 
calibration and testing purposes \cite{GravitySpy}. The detection CNN was able to classify all chirps as BBH signals.}. 
The outcome of this experiment indicates that the glitches need to be included explicitly in the training procedure 
of neural networks intended for realistic GW searches. This extension is left to a subsequent work. We will also 
investigate the effect of the presence of glitches on the GW inference question.

\subsection{Computational efficiency}

Both the detection and regression CNNs are 83 MB in size each and encode information for $\sim150,000$ templates, or 
$\sim45$ GB of data, representing a typical data set used for the neural networks training. In this respect, a 
trained neural network model can be viewed as an abstract and compact representation of the template bank. Most
importantly, since the computationally intensive training stage is performed only once offline, after it has been 
trained, the evaluation of the neural network on new GW data does not depend on the size of training data set.
Therefore, expanding the template bank used for the neural network training to additional dimensions of the signal 
manifold does not affect the processing time. Once trained, processing 10 seconds of GW data takes only $\sim$3 
milliseconds on GPUs with both CNNs. This implies that processing a month of GW data takes about 11 hours (assuming
$\sim10^7$ overlapping GW time series segments of duration 0.3 seconds). Such rapid processing is important for
generating real-time alerts and can provide useful hints for EM counterpart searches and also for focused
data analysis with accurate matched filtering and Bayesian inference approaches \cite{Gebhard:2019ldz}.
The fast speed with which neural networks process data is also advantageous when processing data streams
from multiple GW detectors. In particular, adding more observatories to the detector network does not
increase the computational cost of the deep learning approach as the GW data streams from each detector
(LIGO, VIRGO, KAGRA, etc.) can be processed simultaneously in parallel. Analyzing data from multiple detectors
is even beneficial to the method since applying the detection CNN simultaneously to each data stream, and enforcing
coincidence, eliminates many of the false positives and decreases the FAR \cite{Wei:2020ztw}.  

For instance, as more GW detectors come online, the computational cost of matched filtering methods increases
at least linearly in the number of detectors (because the search for triggers is performed first independently
for each GW detector). In addition, the computational cost, for both trigger generation and Bayesian inference,
scales linearly with increasing the number of waveform templates in the template banks. As template banks
become bigger, matched filtering searches and Bayesian parameter estimation become increasingly computationally 
expensive, which makes online real-time trigger generation and inference extremely challenging \cite{Gebhard:2019ldz}.
In particular, extending the conventional real-time matched filtering approaches to the full 9D signal manifold 
currently available to the GW detectors, and performing real-time conventional Bayesian parameter estimation of GWs, 
is computationally prohibitive at the present \cite{Huerta:2019rtg}. These considerations are the major motivation
for exploring alternative detection and inference methods in the first place.

\section{Summary and outlook}

We have demonstrated for the first time the detection and parameter estimation of GW signals from 
BNS mergers in real advanced LIGO data using deep learning approaches. Specifically, we have shown that deep 
neural networks can recover all GW events from GWTC-1, and rapidly estimate the component masses of simulated BNS
signals in real LIGO noise. These results pave the way towards realizing real-time detection and parameter estimation
of GW signals involving neutron stars, where rapid follow-ups of the EM counterparts are critical. Our findings
also emphasize the importance of using realistic GW detector noise for the training of neural networks intended for 
detection of actual GW events.

The characteristic scalability of deep learning algorithms could enable GW searches covering the full parameter space 
available to GW detectors, and also rapid parameter estimation, which are unfeasible with conventional match-filtering and 
Bayesian inference approaches. Furthermore, multiple neural network instances could take simultaneously GW data streams 
from multiple GW detectors thus enabling consistent GW searches and inference.

Future directions include extending deep learning algorithms to enable classification of GW signals and realistic detector 
glitches in a unified framework, using machine learning approaches for real-time parameter estimation of GW signals from 
BNS and BHNS mergers, and implementing Bayesian neural networks for evaluating error uncertainties. For instance, deep
learning methods could help to deduce key source parameters, such as the neutron star tidal deformability \cite{Hinderer:2009ca},
which is critical for understanding the EOS of dense matter and fundamental inter-particle interactions, but is
theoretically controversial and observationally challenging to measure.

\section*{Acknowledgments}
This work is supported in part by the National Science Foundation under Cooperative Agreement PHY-2019786 (The NSF AI Institute for Artificial Intelligence and Fundamental Interactions, https://iai{f}i.org).
This research has made use of data, software and/or web tools obtained from the Gravitational Wave Open Science Center 
(https://www.gw-openscience.org), a service of LIGO Laboratory, the LIGO Scientific Collaboration and the Virgo Collaboration. 
LIGO is funded by the U.S. National Science Foundation. Virgo is funded by the French Centre National de Recherche Scientifique 
(CNRS), the Italian Istituto Nazionale della Fisica Nucleare (INFN) and the Dutch Nikhef, with contributions by Polish and 
Hungarian institutes. The computational resources were provided by the Faculty of Arts and Sciences Research Computing at 
Harvard University.

\end{document}